\documentclass{llncs}
\usepackage{llncsdoc}

\usepackage{epsfig}
\usepackage{amssymb}
\usepackage{amsmath}
\usepackage{amsfonts}

\usepackage{hyperref}
\usepackage{url}
\usepackage{comment}
\usepackage{float}
\usepackage{amsmath}
\usepackage{amssymb}
\usepackage{algorithm}

\usepackage[noend]{algpseudocode}
\usepackage{caption}

\usepackage{graphicx}
\usepackage[justification=centering]{caption}
\usepackage{multirow}
\usepackage{wrapfig}
\usepackage{dsfont}

\usepackage{subcaption}
\newtheorem{prop}{Proposition}
\begin{document}


 



%


\title{Chiron: A Robust Recommendation System with Graph Regularizer}

\author{Saber Shokat Fadaee \inst{1}  \and Mohammad Sajjad Ghaemi \inst{2} \and Ravi Sundaram \inst{3} \and Hossein Azari Soufiani \inst{4}}
\institute{College of Computer and Information Science\\
Northeastern University \\
email: \url{saber@ccs.neu.edu}
  \and Polytechnique Montréal  \\
email: \url{mohammad-sajjad.ghaemi@polymtl.ca} \and College of Computer and Information Science\\
Northeastern University \\
email: \url{koods@ccs.neu.edu}
  \and Columbia Business School\\
email: \url{Hazarisoufiani18@gsb.columbia.edu}}
\maketitle

\begin{abstract} 
Recommendation systems have been widely used by commercial service providers for giving suggestions to users. Collaborative filtering (CF) systems, one of the most popular recommendation systems, utilize the history of behaviors of the aggregate user-base to provide individual recommendations and  are effective when almost all users faithfully express their opinions. However, they are vulnerable to malicious users biasing their inputs in order to change the overall ratings of a specific group of items. CF systems largely fall into two categories - neighborhood-based and (matrix) factorization-based - and the presence of adversarial input can influence recommendations in both categories, leading to instabilities in estimation and prediction. 
Although the robustness of different collaborative filtering algorithms has been extensively studied, designing an efficient system that is immune to manipulation remains a significant challenge. In this work we propose a novel \emph{hybrid} recommendation system with an adaptive graph-based user/item similarity-regularization - {\bf Chiron}. Chiron ties the performance benefits of dimensionality reduction (through factorization) with the advantage of neighborhood clustering (through regularization). We demonstrate, using extensive comparative experiments, that Chiron is resistant to manipulation by large and lethal attacks.  
\end{abstract} 

\section{Introduction}
\label{sec:intro}

Users of commercial service providers such as Netflix, Spotify, and Amazon are provided with a large selection of recommended choices while using these online services. Recommendation systems aid users in the challenging task of finding the best video, music, book, or product out of all the possible options that they can have while using these systems. In this regard, collaborative filtering-based recommendation systems play an increasing role in helping people locate their favorite items in an immense dataset. In addition to providing helpful recommendations to users, these systems are also beneficial for the companies in raising their sales. However, since a good recommendation usually results in increased sales, some might find it profitable to shill recommendation systems by providing false information.

``Collaborative filtering (CF)'' algorithms predict how much a user  prefers a set of items, and produce a ranked list of items that would benefit or match her interests the most. In recommendation systems based on collaborative filtering, users rate specific items and receive recommendations for unrated ones.   All of these different systems are vulnerable to malicious attackers intending to manipulate the recommendations to suit their needs. Such attackers are known as ``shills'' and those attacks have been referred to as ``shilling'' or ``Sybil'' attacks \cite{Resnick:2008:MRS:1486877.1486887}.

Collaborative filtering methods do not use any information about users or items except for a partially observed rating matrix. The latter contains information provided by different users regarding  different items, and the entries of this matrix are usually either binary or ordinal.  Two of the most popular methods for predicting the missing values are neighborhood based methods: item based CF and user based CF, and matrix factorization: singular value decomposition (SVD) \cite{citeulike:4563135}, Enhanced SVD \cite{Guan2016}, non-negative matrix factorization (NMF) \cite{NIPS2000_1861}, \cite{lee99}, probabilistic matrix factorization (PMF) \cite{NIPS2007_1007}, and Bayesian probabilistic matrix factorization (BPMF) \cite{Salakhutdinov:2008:BPM:1390156.1390267}. There are hybrid recommendation systems that combine both methods.




In this work, we introduce Chiron\footnote{Chiron was the most important Centaur in Greek mythology, and centaurs are hybrid creatures. Since our model is a hybrid-recommendation system that factorizes the user/item matrix and uses the neighborhood information, we picked this name.}, a  robust recommendation system. We conduct extensive experimental studies to compare the robustness of our algorithm with current state-of-the-art methods. Our experimental results indicate that Chiron is the most robust recommendation system, and the presence of an attack does not affect its performance. While many collaborative filtering methods are prone to overfitting, we prevent over-fitting by introducing a smart regularization technique which takes users' and items' similarities into account in the context of local graph estimation of the marginal probability of users and items.
 
\subsection{Related Work}
\label{sec:related_work}

Many psychological studies have shown that people tend to agree with opinions of others regardless of their factual correctness. Cosley et al. \cite{Cosley:2003:SBR:642611.642713} showed that prediction manipulation in a recommendation system can affect people in that system and, in some cases, mislead people into accepting a negative and unfitting recommendation. Therefore, people's perceived value of items are influenced by the ratings of a recommendation system. Chirita et al. \cite{Chirita:2005:PSA:1097047.1097061} demonstrated that the presence of only three attackers in the neighborhood of one user is enough to  create a significant change in prediction and move an unpopular item to the top five recommended items.

Lam et al. \cite{Lam:2004:SRS:988672.988726} and O'Mahony et al. \cite{omahony2002} showed that many of the well-known recommendation systems are vulnerable to attacks and proposed different methods to distinguish honest raters from attackers. 



However, using detection algorithms as a preprocessing step can be computationally expensive. Therefore, others have proposed building robust recommendation systems \cite{Mehta:2008:ARC:1390334.1390350,Mobasher:2007:TTR:1278366.1278372}. While recommendation systems have been widely investigated, less attention has been devoted to studying their vulnerability to manipulations. Mehta et al. \cite{journals/debu/MehtaH08} discussed a robust recommendation system's characteristics. In \cite{citeulike:14072193}, they investigated how different statistical models use locality in order to enforce robustness.

O'Mahony et al. \cite{OMahony06detectingnoise} performed empirical studies on the robustness of user-user kNN algorithms. They showed that attackers can successfully manipulate  recommendation systems both by pushing and nuking attacks. Push attacks happen when malicious users try to increase ratings of specific items, and nuke attacks happen when they do the opposite. 


Seminario  et al. \cite{conf/flairs/SeminarioW12} examined the trade-offs between accuracy and robustness of user-based and item-based CF recommendation systems and showed that the former achieve relatively positive marks on both  properties. However, in an item-based CF recommendation system there exists a trade-off between its accuracy and robustness.


In this work, we propose a new generative model for recommendation systems that not only considers the users' evaluations of items, but also takes into account the items' evaluations of users. In some cases, items could be informative, and have their own evaluations of users. This extra information could be used to add priors to the system in order to promote users who give honest ratings. We incorporate these  evaluations as regularization terms inspired by the local similarities underlying the graph structure.

%
\section{Contribution}
\label{sec:contribution}
We propose a new model subject to certain constraints which considers the quality of a rater (e.g. customers providing feedback), as well as the quality of an item (e.g. a restaurant or a product). Then we apply the neighborhood information using a graph regularizer to approximate a geometrical structure of the distribution of users and items in the latent space.
 
 Fitting an arbitrary model to the observed data is usually prone to over-fitting by increasing the variance of the error term. In order to prevent a model from over-fitting, researchers deploy several techniques, including cross validation and regularization.  Regularization deals with a trade-off between bias and variance of an estimator. It has to ensure a model is complex enough to encompass the observed data smoothly, simultaneously, keeping the model as simple as possible in order to generalize the unobserved data. We have two regularization terms  in our model (\ref{eq:lkreg}). One controlling the user similarity in the intrinsic space of their underlying rated items . The other tunes the item’s rating measured by the geometry of the users' distribution.

 The intuition behind this model is simple yet efficient. Whether someone likes an item or not depends on the affinity between her latent preferences and the item's latent attributes (demonstrated by $\theta$ in our model), and the influence of her friends on her choices and her decision regarding similar items in the past. We add the extra information regarding a person and an item local neighborhood to this model by adding regularization to the log likelihood of our model. Our research is inspired by the work of Belkin et al. \cite{Belkin:2006:MRG:1248547.1248632}, Zheng et al. \cite{Zheng2016}, and Cai et al.\cite{Cai11GNMF}.

We construct our model by first computing a graph from the observed ratings and creating an affinity matrix by measuring the similarities between ratings of different users. Then we incorporate a regularization term into the model which is sufficiently smooth with respect to the intrinsic structure collectively revealed by both observed and missing data. Our model has two basic assumptions:
\begin{enumerate}
\item Local assumption: nearby users are likely to have the same rating.
\item Global assumption: users on the same structure (considered as an underlying manifold) are more likely to have the same rating.
\end{enumerate}
 

The relationship between users can be thought of as an undirected weighted graph, in which the weights reflect the affinity between the ratings of those users. Our regularization propagates the rating's value through the edges attached to it. The value transferred to each user is proportional to its weight (similar to the graph construction in \cite{NIPS2003_2506}).

 
 
 Exploiting the inherent geometry of the marginal distribution could be troublesome if we do not have the true density function\cite{Belkin:2006:MRG:1248547.1248632}. Therefore researchers use transductive learning via spectral graph-Laplacian, which incorporates users and their ratings by extracting the underlying geometric structure to approximate the data-dependent regularizer for this model \cite{NIPS2003_2506}. This is a well studied approach that utilizes both labeled and unlabeled data to improve classification accuracy. \cite{Joachims03transductivelearning}.

We demonstrate how this approach is robust to attacks by malicious users. Additionally, we examine our model's performance on three real-world datasets, and compare the success rate of attacks on our system as well as the state-of-the-art collaborative filtering recommendation systems.  We also compare the running time of different methods on a different dataset and conclude that Chiron is not only robust to attacks, but is also the second fastest algorithm among the varied collection of current recommendation systems.

\section{Model}
\label{sec:model}
Suppose we have $m$ raters who rate $n$ items with a score in the range of $1$ to $K$. Let $P_i = (p_{i1},..,p_{iK})$ be the vector of probabilities of different ratings for item $i$, and $Q_j = (q_{j1},..,q_{jK})$ be the vector of probabilities of different ratings by rater $j$. From now on we use index $j$ to refer to a user, and index $i$ to refer to an item in our model. Let $r_{ij} = k$ if user $j$ gives rating of $k$ to item $i$ and let $z_{ijk}$ denote the corresponding indicator variable, i.e. $z_{ijk} = 1$ if $r_{ij} = k$, and  $z_{ijk} = 0$ otherwise. Now in Chiron, the probability that an item $i$ receives rating $k$ by user $j$ is calculated as follows:

\begin{align}
\label{main_formula1}
Pr(r_{ij} = k) \approx \theta_{ijk} = \frac{p_{ik}q_{jk}}{\sum_{l=1}^K p_{il}q_{jl} } 
\end{align}

\begin{figure}[H]
\centering
\includegraphics[width=0.55\linewidth]{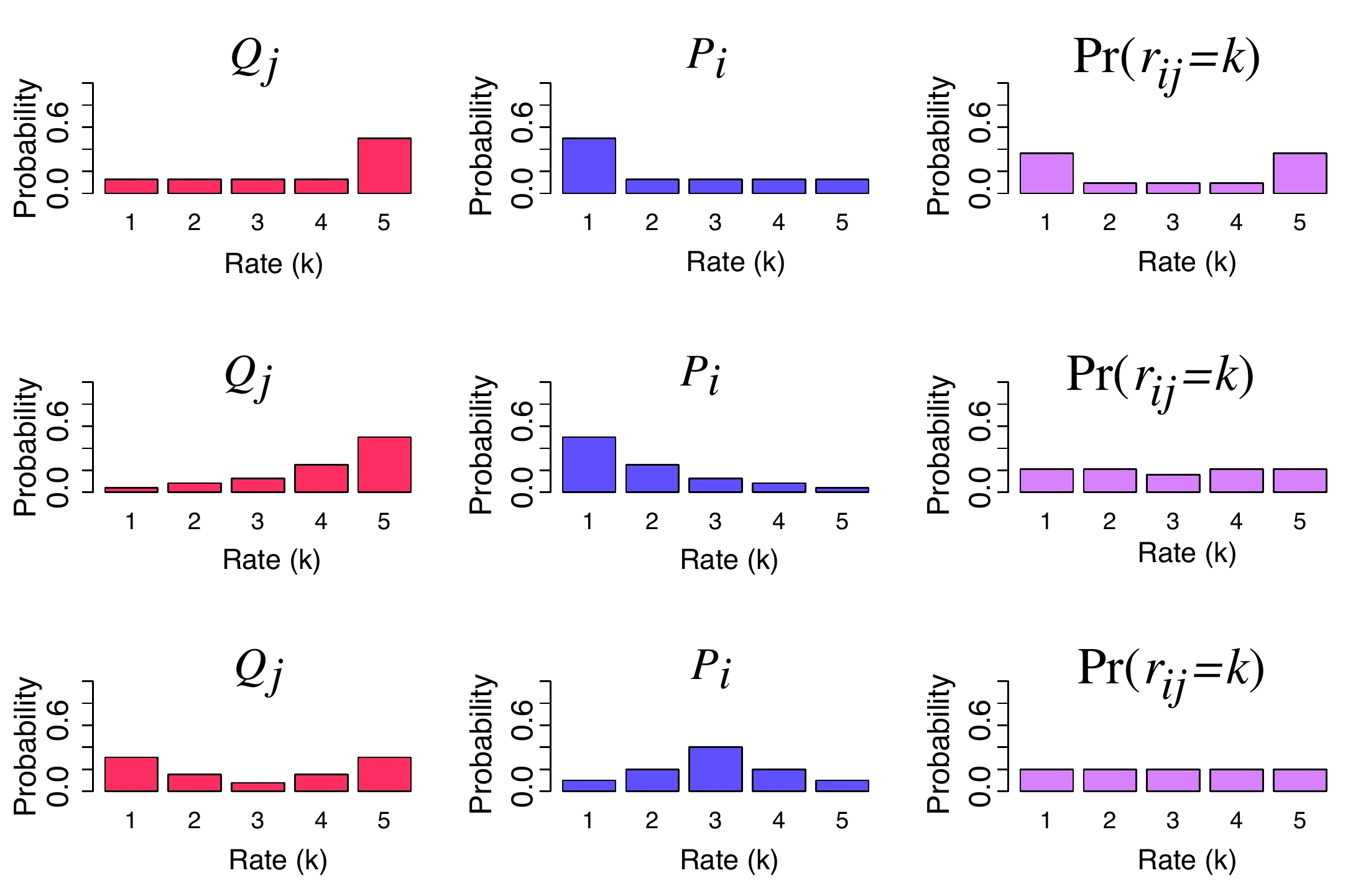}
\caption{Examples of different combinations of probabilities of  ratings given by user $j$  (as $Q_j$) and her ratings for an item $i$ (as $P_i$), and the corresponding probability that item $i$ is rated as $k$ by user $j$ as $Pr(r_{ij} = k)$. }
\label{fig:PQ}
\end{figure}

Different combinations of $P_i$, $Q_j$, and corresponding  $Pr(r_{ij} = k)$ for three cases are illustrated in Figure \ref{fig:PQ}. For example in the first case, we have a user $j$ who gives a high rating to most items (e.g. rates items only when she loves them). For an item $i$ which receives a poor rating by most users and other ratings with the same probability, Chiron predicts that user $j$ gives either the lowest or the highest rating to that item with a high probability. 

In the second case, user $j$ rates items as 1 to 5 with an ascending probability, and item $i$ is rated with a descending probability. In this case, Chiron predicts that this user would rate this item as 1,2,4, or 5 with the same probability and rates it as 3 with a lower probability.


In the last case, user $j$ tends to rate items as either very good or very bad. On the other hand, we have an item $i$ which is mostly rated as average (3). Chiron predicts that user $j$ gives any rating to item $j$ with the same probability. 

$\theta$ captures users' rating habits.  The log likelihood function of \eqref{main_formula1} is: 

\begin{align}
 \mathcal{L}(P,Q;Z) &= \sum_{ijk} z_{ijk}\log \frac{p_{ik}q_{jk}}{\sum_{l}p_{il}q_{jl}} \label{eq:LL} \\ 
&=\sum_{ijk} z_{ijk}\log p_{ik}q_{jk} - \sum_{ij} \log \sum_l p_{il}q_{jl} \label{eq:loglikelihood}   
\end{align}


\section{Estimation}



The model defined in \ref{eq:loglikelihood} is our fitting constraint, which means an appropriate estimating function should not change too much from the initial observed values. Now we add the smoothness constraints, which imply that a reasonable estimating function should not change too much between nearby users and corresponding ratings. The trade-off between these two competing constraints is captured by a positive parameter $\lambda_1$, and its counterpart $1 - \lambda_1$ to make a convex combination of two regularization terms. This way, one can increase the influence of users so that the effect of items will decrease.  We set the value of hyper-parameters $\lambda_1$ using cross-validation.   

\begin{align}
\mathcal{L'}(P,Q;Z) &=  \sum^{n}_{i=1}\sum^{m}_{j=1}  \sum^{K}_{k=1} \left(z_{ijk}\log \frac{p_{ik}q_{jk}}{\sum_{l}p_{il}q_{jl}}  \nonumber \right. \\ 
	&+\lambda_1 q_{jk} (d^{Q}_{jj}q_{jk}-\sum^{m}_{t=1}q_{tk}w^{Q}_{tj}) 	\nonumber \\ & \left.+ (1- \lambda_1) p_{ik} (d^{P}_{ii}p_{ik}-\sum^{n}_{t=1}p_{tk}w^{P}_{ti})\right)
\label{eq:lkreg}
\end{align}

In formula \ref{eq:lkreg} $w^{Q}$ and $w^{P}$ represent the graph underlying the data for users and items respectively. We represent those graphs with $G^{Q} = (V^{Q},E^{Q})$, and $G^{P} = (V^{P},E^{P})$. In $G^{Q}$ and $G^{P}$ users or items with similar ratings are connected to each other. To construct the weight matrix $w^{Q}$ (we use similar method for computing $w^{P}$), we find $10$ nearest users for each user in $Q$ using Pearson correlation (to construct $E^{Q}$). Then we apply the following kernel for each connected pair $(j_{1}, j_{2})$: 
\[
w^{Q}_{j_{1} j_{2}}=\sum^{n}_{i=1}\sum^{K}_{k=1}z_{ij_{1}k}z_{ij_{2}k}
\]
$w^{Q}_{j_{1}j_{2}}$ measures how many times both users $j_{1}$ and $j_{2}$ give the same rating to the same item. 


We use a regularization term that keeps the model flexible enough to assign different probabilities to $q_{j_{1}k}$, and $q_{j_{2}k}$ for distant users $j_{1}$, and $j_{2}$ in order to keep them away from each other. This difference is proportional to the small weight in terms of rating similarity that connects them weakly. The regularization term $w^{Q}_{j_{1}j_{2}}(q_{j_{1}k}-q_{j_{2}k})^2$ best captures this property, which means it pushes the $(q_{j_{1}k}-q_{j_{2}k})$ toward zero for strong similarity between $(j_{1}, j_{2})$ pair with similar rating.


Using local assumption that nearby users have similar ratings, the discrete $k=10$-nn graph estimates the global manifold for the underlying users. The observed ratings can propagate through this estimated graph to impute the missing values. Consequently, in order to spread the information among users, Laplacian of the constructed graph plays the central role in predicting the missing data from the observed values. Laplacian denoted by $L^{Q}=D^{Q}-W^{Q}$, is considered in this study where $D^{Q}$ is a diagonal matrix whose entries are $d^{Q}_{jj}=\sum_{t}w^{Q}_{jt}$. $L^{Q}$ is a symmetric and positive semi-definite matrix. This representation allows the information of ratings propagates smoothly between users with probability that is proportional to the weight between them. For the sake of calculating the stochastic element-wise gradient, we expand the $w^{Q}_{j_{1}j_{2}}\sum^{K}_{k=1}(q_{j_{1}k}-q_{j_{2}k})^2$ term within the likelihood cost function as the sum element-wise notation expressed in equation~\ref{eq:lkreg}.

The model defined in \ref{eq:loglikelihood} minimizes the variance term for the true estimator $\theta$. The class of unconstrained models are usually prone to overfitting. At the same time simple models suffer from underfitting due to increase in bias. We suggest a smart regularization method based on locality preservation to compromise both issues. Hence, we consider a specific assumption regarding the connection between the marginal and the conditional distributions of $P$ and $Q$. Let us assume that if two items $p_{1}, p_{2} \in P$ share close ratings in the form of intrinsic geometry of $Pr(P)$, then this implies the conditional distributions $Pr(Q|p_{1})$ and $Pr(Q|p_{2})$ are correspondingly alike. In other words, the conditional probability distribution $Pr(Q|p_{i})$ varies smoothly along the geodesics in the true geometrical shape of $Pr(P)$. Therefore, we have two regularization terms, one controls user similarities in the intrinsic space of their underlying rated items. The other tunes  rating of items using the geometry of users distribution.

In order to compute the maximum likelihood of our model, we take the partial derivative of equation \eqref{eq:lkreg} with respect to \textit{P} and \textit{Q}:

\begin{align}
\frac{\partial \mathcal{L}(P,Q;Z) }{\partial q_{j^{*}k^{*}}} &=  \sum_{i} z_{ij^*k^*} \frac{p_{ik^*}}{q_{j^*k^*}} - \sum_{ij} \frac{p_{ik^*}}{\sum_l p_{il}q_{jl}} +\lambda_1  (d^{Q}_{j^*j^*}q_{j^*k^*}-\sum^{m}_{t=1}q_{tk^*}w^{Q}_{tj^*}) \label{eq:derivative1} \\
\frac{\partial \mathcal{L}(P,Q;Z) }{\partial p_{i^{*}k^{*}}} &=  \sum_{j} z_{i^*jk^*} \frac{q_{jk^*}}{p_{i^*k^*}} - \sum_{ij} \frac{q_{jk^*}}{\sum_l p_{il}q_{jl}}+ (1 - \lambda_1)  (d^{P}_{i^*i^*}p_{i^*k^*}-\sum^{n}_{t=1}p_{tk^*}w^{P}_{ti^*})
\label{eq:derivative}
\end{align}

And then we set them to zero which leads to the following equations:

\begin{align}
q_{j^{*}k^{*}} &= \frac{\sum_{i} z_{ij^{*}k^{*}} p_{ik^*}}{\sum_{i} \frac{p_{ik^{*}}}{\sum_{l}p_{il}q_{j^{*}l}} z_{ij^{*}k^*}-\lambda_1  (d^{Q}_{j^*j^*}q_{j^*k^*}-\sum^{m}_{t=1}q_{tk^*}w^{Q}_{tj^*})} 
\label{eq:updQ}  \\
p_{i^{*}k^{*}} &= \frac{\sum_{j} z_{i^{*}jk^{*}}q_{jk^*}}{\sum_{j} \frac{q_{jk^{*}}}{\sum_{l}p_{i^{*}l}q_{jl}} z_{i^{*}jk^*} - (1 - \lambda_1)  (d^{P}_{i^*i^*}p_{i^*k^*}-\sum^{n}_{t=1}p_{tk^*}w^{P}_{ti^*})} 
\label{eq:updP}
\end{align}

Since we would like to optimize both $p_{ik}$ and $q_{jk}$, our model is bi-convex that is prone to get trapped in one of the local optimums. One possible solution is to fix one of the unknown parameters, and solve the optimization problem for the other. We use the average alternating projections
method \cite{DBLP:journals/corr/AgarwalA0NT13} with different initial values to provide a set of estimators.  For this purpose we first fix  $p_{ik}$ and solve the optimization problem for  $q_{jk}$, and then fix  $q_{jk}$ and solve the problem for  $p_{ik}$, and continue until convergence. We assume the model converges when the following holds for Q (and  a similar term for P):
\begin{align}
\frac{1}{m}\sum_{1 \leq j \leq m} \sum_{k} (q_{j k} - q_{j*k*})^2 < \epsilon
\label{eq:conv1}
\end{align}

In \eqref{eq:conv1}  $\epsilon$ is a small number ($10^{-3}$ in our case). In the two real-world data sets that we examined, Chiron converges in at most 5 steps. 

\section{Experimental Setting}
\label{sec:design}

We examined the robustness, and accuracy of Chiron in rating prediction and compared it with state-of-the-art recommendation systems.

In subsection \ref{subsec:dataset} we introduce the datasets we are using in our experiments. Then in \ref{subsec:attack-design} we introduce various attack strategies, and evaluation methods for comparing shilling attacks. After that we explain why we chose a specific kind of attack for our experiments. Finally in \ref{subsec:experiments}, we introduce the different recommendation systems, and the toolkit we are using to compare them.

\subsection{Datasets}
\label{subsec:dataset}

We used the MoveLens100K, and Netflix3m1k databases for our experiments. The first data set is gathered by GroupLens Research Project \cite{movielens} at the University of Minnesota. The last one is provided by Netflix in the Netflix prize \cite{Bennett07thenetflix}. Prea  software \cite{Lee:2012:PPR:2503308.2503328} gathered all these data sets in its toolkit. 
In each dataset each user rated at least 20 movies from 1 (defined as did not like) to 5 (liked very much). We performed a cross-validation by splitting each of dataset into a training set ($80\%$), a validation set ($10\%$), and a test set ($10\%$), and compared the predicted ratings with actual ratings of the test set. We repeated our experiments 10 times and used the averaged results.

\subsection{Attack Design}
\label{subsec:attack-design}


In this paper, we are only concerned with shilling attacks in which attackers try to manipulate a recommendation system by introducing fake users, and subsequently fake ratings. We only focus on push attacks since they are usually more successful than nuke attacks \cite{Lam:2004:SRS:988672.988726}. The effect of an attack is measured by the deviation in predicted ratings before and after adding the attack profiles. The most common metric for evaluating recommendation systems is Mean Absolute Error(MAE) which is used to measure accuracy in predicting ratings.

Two important metrics that are used for evaluation of different shilling attacks are the attack size and the filler size \cite{journals/debu/MehtaH08}.The attack size is the ratio of added shilling profiles to the original data set. For example, a $10\%$ attack size indicates that the number of shilling profiles added to the system is equal to $10\%$ of the users in the original data set. Another metric that is used for evaluation of different shilling attacks is the filler size. The filler size is the set of items which are voted for in the attacker profile.

We target a set of 20 items for the push attack. We  repeat each experiment 10 times, and consider the mean value across these 10 times for each item in order to make sure our results are statistically significant.

The most effective attack models are derived by reverse engineering the recommendation algorithms to maximize their impact. As Burke  et al. \cite{Burke:2006:CFA:1150402.1150465} mentioned, the most common recommendation systems attack methods are random, average, and bandwagon. In a random attack, the assigned ratings made by attackers are around the overall mean rating with standard deviation 1.1. In average attacks, the assigned ratings made by attackers are around the mean rating of every item and standard deviation 1.1. Bandwagon attack is similar to the random attack, and some popular items are rated with the maximum rate.


Random and Bandwagon attacks do not require much knowledge about the set of items they are attacking. They only need information about some popular items and their overall means. Creating random ratings within a certain average interval will allow the attacker to have a high impact in making decisions for other users. On the other hand, average attacks require more information and are shown to be near optimal in impact \cite{Mehta:2007:USD:1619797.1619870}. They are also very challenging to detect \cite{Zhang:2006:ALL:1148170.1148259},  and are stronger than random or bandwagon attacks \cite{journals/debu/MehtaH08}. Therefore, in this work we are only concerned with the average attacks. 


\subsection{Experiments}
\label{subsec:experiments}

We use Prea \cite{2012arXiv1205.3193L} to compare our proposed model with different recommendation systems. The different algorithms we select to compare with Chiron fall into these two categories: memory-based neighborhood methods, and matrix factorization methods.

Memory-based neighborhood methods use the knowledge about similar users or items to give predictions about the unrated items. Memory based methods that we use in this experiment for comparison are: User-based Collaborative Filtering, User-based Collaborative Filtering (Default Voting), Item-based Collaborative Filtering, Item-based Collaborative Filtering (Inverse User Frequency), and Slope One.

On the other hand, matrix factorization methods build low-rank user or item profiles by factorizing training datasets with linear algebraic methods.The matrix factorization methods that we use in this experiment are: Regularized SVD, Non-negative Matrix Factorization (NMF), Probabilistic Matrix Factorization (PMF),  and Bayesian Probabilistic Matrix Factorization (BPMF).

\section{Results and discussion}
\label{sec:results}

The results of running different recommendation systems on the  Netflix3M1K, and MovieLens100K, are shown in Table \ref{table1}. We compared the prediction accuracy growth among different collaborative system methods after an attack size $100\%$ on the first two data sets. A $100\%$ attack is a shill, in which the number of fake users is equal to the number of genuine users. The mission of attackers is to promote a certain list of items (20 items in our experiment) and give an average rating to another set of random items to remain undetected. As expected, Chiron is the most robust recommendation system and has the least amount of change in accuracy.


\begin{table*}[t]
\centering
\small{

\begin{tabular} {| c | c |c |c | c | c| c | c | c | c |}
\hline
 \multicolumn{1}{|c|}{}&\multicolumn{6}{c|}{Data set}\\ 
 \cline{2-7} 
 \multicolumn{1}{|c|}{Method}&\multicolumn{3}{c|}{Netflix3M1K}&\multicolumn{3}{c|}{MovieLens100k}
 \\
 \cline{2-7} 
 \multicolumn{1}{|c|}{}&\multicolumn{6}{c|}{MAE}\\ 
 \cline{2-7} 
	& Before	& After  &	 Growth & Before	& After &Growth   \\ \hline
\hline 
user-based CF & 0.772 & 0.985 & $27\%$& 0.734 & 0.924 & $25\%$ \\ \hline
user-based DF & 0.765 & 0.979 & $27\%$& 0.735 & 0.922 & $25\%$  \\ \hline
item-based CF & $\mathbf{0.756}$ & 0.970 &	$28\%$& 0.722 & 0.923 & $27\%$  \\ \hline 
item-based DF & 0.760 & 0.979 & $28\%$& $\mathbf{0.718}$ & 0.923 & $28\%$ \\ \hline 
Slope One	& 0.775   & 1.045 & $34\%$& 0.744 & 0.985 & $32\%$ \\ \hline \hline
Regular SVD	& 0.819   & 1.528 & $86\%$& 0.729 & 0.982 & $34\%$ \\ \hline
Non negative MF &0.868 & 1.745&$101\%$& 0.780 & 1.043 & $33\%$  \\ \hline
Probabilistic MF &0.786 & 1.280&$62\%$& 0.775 & 0.984 & $26\%$ \\ \hline
Bayesian PMF&0.793 & 1.319 & $66\%$& 0.745 & 0.977 & $31\%$ \\ \hline
Chiron & 0.775 & $\mathbf{0.943}$ & $\mathbf{21\%}$& 0.737 &$\mathbf{0.908}$ & $\mathbf{23\%}$ \\ \hline
\end{tabular}
}
\caption{Changes in prediction accuracy after an attack size of $100\%$ in the Netflix3M1K, and MovieLens100k data sets. The statistically significant result is shown in bold in each column.}
\label{table1}
\end{table*}

\begin{figure*}[t!]
    \centering

\captionsetup{justification=centering}

\begin{subfigure}[b]{0.45\textwidth}
\centering
\includegraphics[width=1.0\linewidth]{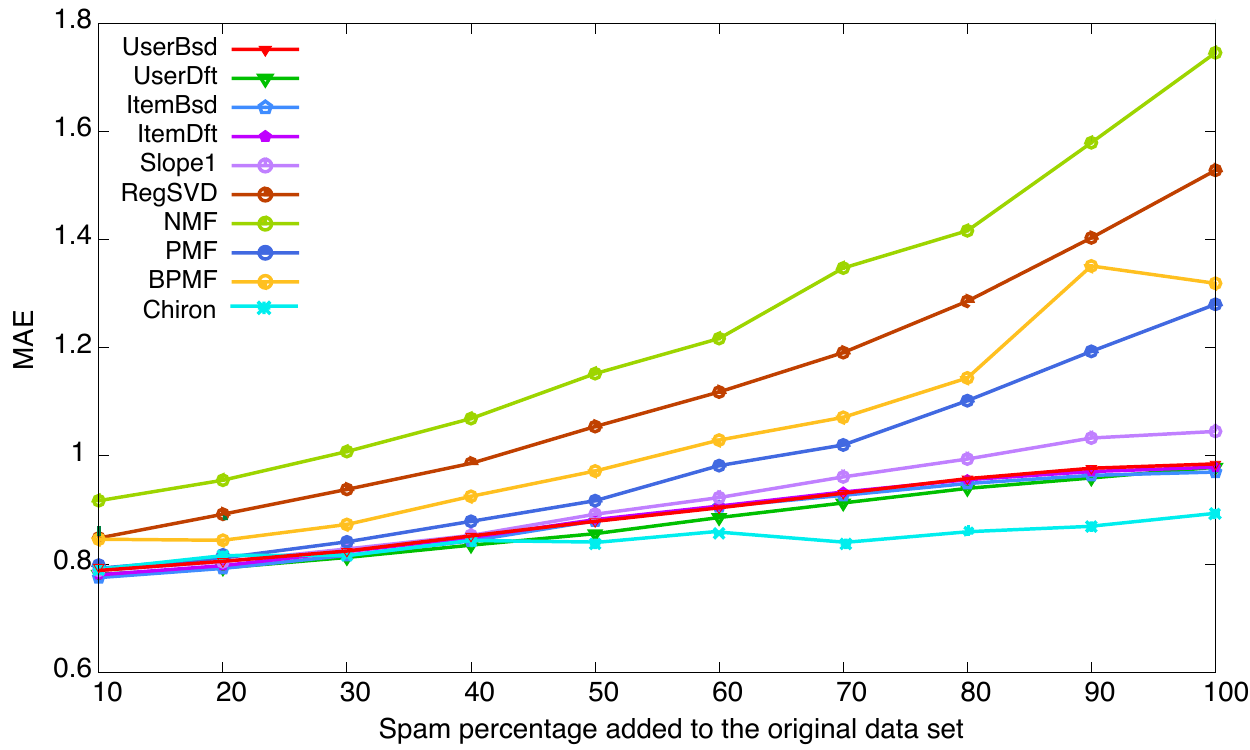}
\caption{Changes in prediction accuracy \\ Data set: Netflix3M1K}
\label{fig:plot2-1}
\end{subfigure}
~
\begin{subfigure}[b]{0.45\textwidth}
\centering
\includegraphics[width=1.0\linewidth]{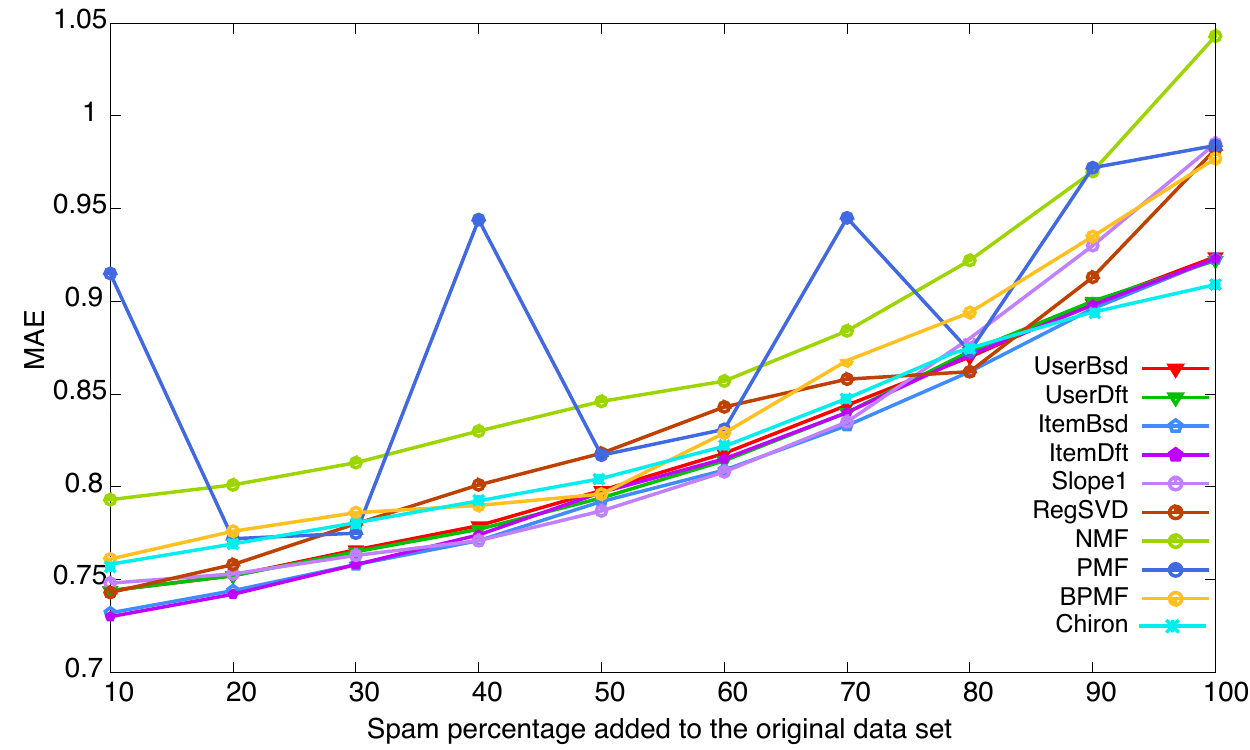}
\caption{Changes in prediction accuracy \\Data set: MovieLens100K}
\label{fig:plot2-2}
\end{subfigure}
\caption{Changes in prediction accuracy in two different data sets.}
\end{figure*}

In another similar experiment, we attacked both the Netflix3M1K and MovieLens100K data sets with a $10\%$ attack and gradually increased the attack size until it reached $100\%$. We increase the size till $100\%$ to demonstrate how different models react to the increase in attack size. We pick $100\%$ because most models could be distinguished from each other at that point. The changes in prediction accuracy are illustrated in Figure \ref{fig:plot2-1} and \ref{fig:plot2-2}. In both of them the performance of Chiron almost remains unchanged with increases in the attack size while the prediction accuracy of other methods drops. 



%

\section{Conclusion}
\label{sec:conclusion}

Various methods exist for protecting recommendation systems against attacks by malicious users. More research has been done in the detection of attackers rather than proposing a robust recommendation system.  Besides, the level of spam in real world data is often high, and simple spam detection methods are often reverse engineered. In this paper, we proposed a new  model which is empirically robust to attacks, explored its characteristics, and provided compelling evidence of its robustness. Chiron can not easily be manipulated since it relies on modeling and other users ratings. Furthermore, we have illustrated its improved robustness in comparison with other state-of-the-art methods, and concluded that Chiron does not lose accuracy in the presence of shill attacks.

\bibliographystyle{plain}

\bibliography{bibliography}

\end{document}